\documentclass[12pt]{article}
\usepackage{epsfig}
\begin{document}
\def\be{\begin{equation}}
\def\ee{\end{equation}}
\def\bq{\begin{equation}}
\def\eq{\end{equation}}
\def\bqa{\begin{eqnarray}}
\def\eqa{\end{eqnarray}}
\def\roughly#1{\mathrel{\raise.3ex
\hbox{$#1$\kern-.75em\lower1ex\hbox{$\sim$}}}}
\def\lsim{\roughly<}
\def\gsim{\roughly>}
\def\llgm{\left\lgroup\matrix}
\def\rrgm{\right\rgroup}
\def\vectrl #1{\buildrel\leftrightarrow \over #1}
\def\partrl{\vectrl{\partial}}
\def\gslash#1{\slash\hspace*{-0.20cm}#1}

\begin{flushright}
BI-TP 2008/07
\end{flushright}

\begin{center}
{\bf The Color Dipole Picture and the ratio of 
$R (W^2, Q^2) = \sigma_L/\sigma_T$}\footnote{Supported by Deutsche 
Forschungsgemeinschaft, contract number schi 189/6-2 and the Ministry 
of Education and Science, Japan under the Grant-in-Aid for basic
research program B (No. 17340085).}\footnote{email:
kurodam@law.meijigakuin.ac.jp; Dieter.Schildknecht@physik.uni-bielefeld.de}
\end{center}
\vspace {0.5 cm}
\begin{center}
{\bf Masaaki Kuroda}$^a$  and
{\bf Dieter Schildknecht$^b$ }\\[2.5mm]
$a$ Institute of Physics, Meiji Gakuin University \\[1.2mm]
Yokohama 244-8539, Japan \\[1.2mm]
$b$ Fakult\"{a}t f\"{u}r Physik, Universit\"{a}t Bielefeld \\[1.2mm] 
D-33501 Bielefeld, Germany \\[1.2mm]
and \\[1.2mm]
Max-Planck Institute f\"ur Physik (Werner-Heisenberg-Institut),\\[1.2mm]
F\"ohringer Ring 6, D-80805, M\"unchen, Germany
\end{center}

\vspace{2 cm}

\begin{center}
{\bf Abstract}
\end{center}

The transverse size of $q \bar q$ fluctuations of the longitudinal
photon is reduced relative to the transverse size of $q \bar q$
fluctuations of the transverse photon. This implies $R (W^2, Q^2) =
0.375$ or, equivalently, $F_L/F_2 = 0.27$ at $x\ll 0.1$ and $Q^2$ 
sufficiently large, while $R (W^2, Q^2) = 0.5$, if
this effect is not taken into account. Forthcoming experimental data from
HERA will allow to test this prediction.

\vfill\eject

In the present paper, we present our prediction for the ratio of the
longitudinal to the transverse photoabsorption cross section, 
$R (W^2, Q^2)
 \equiv \sigma_{\gamma_L^*} (W^2, Q^2)/\sigma_{\gamma_T^*} (W^2, Q^2)$, 
in the diffraction
region of low values of the Bjorken variable $x \cong Q^2/W^2 \ll 1$.
The prediction is based on a careful reconsideration of our previous
formulation \cite{Surrow} of the color-dipole picture (CDP) \cite{Nikolaev}. 

At low values of $x \ll 1$, in terms of the
imaginary part of the virtual forward Compton-scattering amplitude, 
deep inelastic scattering (DIS)
proceeds via forward scattering of (timelike) quark-antiquark, $q \bar q$,
fluctuations of the virtual spacelike photon on the proton. In its interaction
with the proton, a $q \bar q$ fluctuation acts as a color dipole. A
massive $q \bar q$ fluctuation is identical to the $(q \bar q)^{J=1}$ vector
state originating from a timelike photon in $e^+e^-$ annihilation at an
$e^+e^-$ energy equal to the mass, $M_{q \bar q}$, of the $q \bar q$ state.

A well-known life-time argument \cite{lifetime} allows one to put upper
bounds on the magnitude of the scaling variable, $x \cong Q^2/W^2$, and
on the magnitude of the contributing $q \bar q$ masses, $M_{q \bar q}$.
Validity of the color-dipole-picture (CDP) requires the life-time 
of a $q \bar q$ fluctuation
\be
L = \frac{W^2}{Q^2 + M^2_{q \bar q}} \frac{1}{M_p} \equiv L_0 \frac{1}{M_p}
\label{1}
\ee
to be large compared with the scale set by the proton mass, $M_p$.

Requiring
\be
L_0 = \frac{1}{x + \frac{M^2_{q \bar q}}{W^2}} \gg 1
\label{2}
\ee
implies

\begin{itemize}
\item[i)] small x, e.g. $x \ll 0.1$, as well as
\item[ii)] a restriction on the masses of the
$q \bar q$ states that actively contribute to the scattering
process at a given center-of-mass energy $W$,
\be
\frac{M^2_{q \bar q}}{W^2} \ll 0.1 .
\label{3}
\ee
\end{itemize}
Specifying the restriction (\ref{3}) to
\be
     \frac{M^2_{q \bar q}}{W^2} = 0.01,
\label{4}
\ee
for e.g. a center-of-mass energy of W = 225 GeV, we obtain $M_{q \bar q} =
22.5 GeV$. 
The upper bound on the actively contributing dipole
states must coincide with the upper end of the mass spectrum for 
copious diffractive production in $\gamma^*$-proton scattering at
any given energy. Indeed, the upper end of the diffractive mass
spectrum at $W = 225 GeV$ at HERA \cite{HERA} roughly coincides with
the simple estimate based on the crude limit adopted in (\ref{4}). In
most applications \cite{Golec} of the CDP, the upper bound on the 
$q \bar q$ mass is ignored. Such an approximation is valid under
kinematic restrictions on $Q^2$ and $x$, compare ref.
\cite{PRD}. A suitable upper bound  extends \cite{PRD} the kinematic
region of validity of the CDP.

With respect to the ensuing discussions, it will be useful to start
\cite{Cvetic} from the transition of a timelike photon, $\gamma^*$, for
definiteness assumed to originate from an $e^+e^-$-annihilation process,
to a $q \bar q$ pair. The mass of the $q \bar q$ pair, $M_{q \bar q}$, is
identical to the $\gamma^*$ energy in the $q \bar q$ rest frame. A Lorentz
boost leads from the $q \bar q$ rest frame to the proton rest frame.
The direction of the Lorentz boost coincides with the direction
of the $\gamma^*$ three-momentum
in the proton rest frame. The relation between the $q \bar q$ rest frame
and the proton rest frame is assumed to be such that the $\gamma^*p$
center-of-mass energy is much larger than the $q \bar q$ mass,
$W \gg M_{q \bar q}$.

In terms of the transverse momentum, $\vec k_\bot$, of the (massless)
quark and antiquark with respect to the photon direction, the mass
$M_{q \bar q}$ is given by
\be
    M^2_{q \bar q} = \frac{\vec k^{~2}_\bot}{z(1-z)} \equiv 
    \vec k^{~\prime 2}_\bot.
\label{5}
\ee
Here, $0 \le z \le 1$, denotes the usually employed variable that is related
to the $q \bar q$ rest-frame angle between the 
$\gamma^*p$-axis and the three-momentum of the quark,
\be
    \sin^2 \vartheta = 4 z (1-z).
\label{6}
\ee
The electromagnetic
$q \bar q$ current determines the coupling strength of the timelike photon 
to the $q \bar q$ pair
of mass $M_{q \bar q}$. The squares of the longitudinal and the transverse
component  of the electromagnetic $q \bar q$
current with respect to the $\gamma^*p$ axis are given by \cite{Cvetic}
\be
\sum_{\lambda = - \lambda^\prime = \pm 1} \vert j_L^{\lambda, \lambda^\prime}
\vert^2 = 8 M^2_{q \bar q} z (1-z)
\label{7}
\ee
and
\be
\sum_{\lambda = - \lambda^\prime = \pm 1} \vert j_T^{\lambda, \lambda^\prime}
(+) \vert^2 = \sum_{\lambda = - \lambda^\prime = \pm 1} 
j_T^{\lambda, \lambda^\prime} (-) \vert^2 = 2 M^2_{q \bar q} (1 - 2 z (1-z)).
\label{8}
\ee
Here, $\lambda = - \lambda^\prime = \pm 1$ refers to twice the helicity of the 
massless quark and the antiquark, and $j_T^{\lambda, \lambda^\prime} (+)$ and
$j_T^{\lambda, \lambda^\prime} (-)$ refer to positive and negative
helicity, respectively, of the transversely polarized photon. 
Note that $z (1-z)$ in
(\ref{7}) and (\ref{8}) may be replaced by the production angle in the 
$q \bar q$ rest frame according to (\ref{6}).

The $q \bar q$ state of mass $M_{q \bar q}$ originating from the coupling
to the photon consists of a quark and an antiquark of opposite helicity
forming a spin 1 (vector) state. In the limit of very high energy, 
$W \gg M_{q \bar q}$, the longitudinal momenta of the quark and 
antiquark in the proton rest frame, 
in good approximation, become equal in magnitude, independently of the value
of  $0 \le z \le 1$; the $q \bar q$ vector state, as far as the
longitudinal momenta of the quark and antiquark are concerned, does not
contain any memory on the value of $0 \le z \le 1$ it originated from.
The interaction cross section of the $q \bar q$ pair with the proton, as far
as the longitudinal quark and antiquark momenta are concerned is
independent of $z$, and independent on whether  the $q \bar q$ state originates
from a longitudinal or a transverse photon. The magnitude of the
longitudinal momentum components in this high-energy limit only enters via
a dependence of the $q \bar q$-proton interaction on the energy, $W$.

The situation is different with respect to the transverse momentum of the
quark and antiquark. The difference in the transverse momenta for
different values of $z$ at fixed mass, $M_{q \bar q}$,
\be
\vec k^2_\bot = z (1-z) M^2_{q \bar q}
\label{9}
\ee
is independent of the value of $W$, and it remains the same 
specifically also in the high-energy limit of $W \gg M_{q \bar q}$
under consideration in the present context. The normalized distributions
of the quark (antiquark) transverse momentum resulting from the coupling 
strengths in (7) and (8), for longitudinal and transverse photons,
\be
f_L (z) \equiv \frac{z(1-z)}{\int dz~z(1-z)} = 6z(1-z),
\label{10}
\ee
and
\be
f_T (z) \equiv \frac{1-2z(1-z)}{\int dz (1-2z (1-z))} = \frac{3}{2}
(1-2z(1-z)),
\label{11}
\ee
imply different average transverse momenta squared of the quark (antiquark) 
originating from longitudinal and transverse photons,
\be
\langle \vec k^{~2}_\bot \rangle_{L,T} = M^2_{q \bar q} \int^1_0
dz~ z (1-z) f_{L,T} (z).
\label{12}
\ee
Explicitly one obtains from (\ref{12}),
\be
\langle \vec k^{~2}_\bot \rangle_L = \frac{4}{20} M^2_{q \bar q},
\label{13}
\ee
and
\be
\langle \vec k^{~2}_\bot \rangle_T = \frac{3}{20} M^2_{q \bar q},
\label{14}
\ee
and from (\ref{13}) and (\ref{14}),
\be
\rho = \frac{\langle \vec k^{~2}_\bot \rangle_L}{\langle \vec k^{~2}_\bot
\rangle_T} = \frac{4}{3}.
\label{15}
\ee
The result (\ref{15}) is qualitatively expected, since a non-vanishing
transition of a longitudinal photon, $\gamma^*_L$, to a $q \bar q$ pair
according to (\ref{7}) requires $z \not= 0,1$, or equivalently, a 
non-vanishing rest-frame production angle (\ref{6}), in distinction from
the transverse case (\ref{8}), where $z = 0,1$ is by no means excluded.

From the uncertainty relation, the ratio of the effective transverse sizes,
$\langle r^2_\bot \rangle_{L,T}$, for longitudinal and transverse photons
according to (\ref{15}) is given by\footnote
{The $q\bar q$ state originating from a longitudinally polarized 
photon, $\gamma_L^*$, only differs in the average (transverse) 
momentum-squared, $\langle\vec k_\perp^2\rangle_L$, from the $q \bar q$ state 
originating from a transversely polarized photon,
$\gamma_T^*$.  Accordingly, the (transverse) size-squared associated with
$\gamma^*_T$, namely $\langle\vec r_\perp^2\rangle_T$,
is obtained from the (transverse) size-squared associated with $\gamma^*_L$, 
namely $\langle\vec r_\perp^2\rangle_L=A/\langle\vec k_\perp^2\rangle_L$,
where $A\le {1\over 4}$, by the replacement $\langle\vec k_\perp^2\rangle_L
\to \langle\vec k_\perp^2\rangle_T$, i.e. 
$\langle\vec r_\perp^2\rangle_T=A/\langle\vec k_\perp^2\rangle_T$
without change of $A$ under this replacement.  Relation (\ref{16})
follows immediately.}
\be
\frac{\langle \vec r^{~2}_\bot \rangle_L}{\langle \vec r^{~2}_\bot \rangle_T}
= \frac{1}{\rho} = \frac{3}{4}.
\label{16}
\ee
Longitudinal photons, $\gamma^*_L$, produce ``small-size'' pairs, while
transverse photons, $\gamma^*_T$, produce ``large-size'' pairs. The ratio
of the average sizes at any fixed $q \bar q$ mass is given by (\ref{16}).

We summarize: as far as the longitudinal quark momenta are concerned, their
approximate equality in the high-energy limit of $W \gg M_{q \bar q}$
implies that they only affect the $q \bar q$-proton interaction via a 
dependence on $W$ that is independent on whether the $q \bar q$ state
originates from a longitudinal or a transverse photon. In contrast, the
difference in the average quark (antiquark) transverse momenta
(\ref{13}) and (\ref{14}) for longitudinal and transverse photons will
affect the $q \bar q$-proton interaction via the transverse size of the
$q \bar q$ state according to (\ref{16}).

We take the different interaction size into account by introducing 
a proportionality factor, $\rho$, connecting the $q \bar q$-proton 
interactions induced by longitudinal, $(q \bar q)^{J=1}_L$, and transverse,
$(q \bar q)^{J=1}_T$, quark-antiquark states of mass $M_{q \bar q}$,
\be
\sigma_{(q \bar q)^{J=1}_T p} (M^2_{q \bar q}, W^2) = \rho~~
\sigma_{(q \bar q)^{J=1}_L p} (M^2_{q \bar q}, W^2).
\label{17}
\ee
The value of $1/\rho$ from (\ref{16}) suppresses ``small-size'' 
longitudinal versus
``large-size'' transverse hadronic $q \bar q$ cross sections.  Note that 
$\sigma_{(q \bar q)^{J=1}_T} \equiv \frac{1}{2} (\sigma_{(q \bar q)^{J=1}_{+1}}
+ \sigma_{(q \bar q)^{J=1}_{-1}}) = \sigma_{(q \bar q)^{J=1}_{+1}}$, i.e.
$\rho = 1$, instead of (\ref{16}), is identical to helicity independence 
for $(q \bar q)$ vector-state
scattering\footnote{In the talk at DIS2008 given by one of us (D.S.), 
helicity independence was erroneously presented as necessary. The written version of 
the contribution to DIS2008\cite{DIS08} agrees with the results of the 
present paper, see also ref. \cite{DIS07}, where helicity independence is
introduced as a hypothesis.}, $\sigma_{(q\bar q)_{+1}^{j=1}} =
\sigma_{(q\bar q)_{-1}^{j=1}} =\sigma_{(q\bar q)_{0}^{j=1}}$ . 
We stress that the suppression of small-size
longitudinal versus large-size transverse $(q \bar q)^{J=1}$-scattering
cross sections is independent of the mass, $M_{q \bar q}$, of the
$(q \bar q)^{J=1}$ state under consideration. As a consequence, when passing
to $q \bar q$ fluctuations of spacelike photons, the suppression effect
enters as a normalization factor, as in (\ref{17}), while being irrelevant
for the $Q^2$ dependence. This is at variance with the frequently presented
discussion, e.g. ref. \cite{Barone}, where transverse-size effects are
associated with the $Q^2$ dependence of the longitudinal versus the transverse
photoabsorption cross section.

The transition from the scattering of a $q \bar q$ pair of mass
$M_{q \bar q}$ to the scattering of (timelike) $q \bar q$ fluctuations 
of a virtual spacelike photon may be described in transverse position
space in terms of the so-called photon wave function \cite{Nikolaev}.
In terms of the variable $\vec r^{~\prime}_\bot$, related to the transverse
quark-antiquark distance in (\ref{16}) via 
\be
    \vec r^{~\prime}_\bot = \vec r_\bot \sqrt{z (1-z)},
\label{18}
\ee
the longitudinal and the transverse photoabsorption cross section is given by
\cite{PRD}
\bqa
&&   \sigma_{\gamma^*_{L,T}p} (W^2, Q^2)  =  \int dz \int 
   \frac{d^2 r^\prime_\bot}{z(1-z)} \vert \psi_{L,T} (r^\prime_\bot Q, z(1-z)) 
   \vert^2 \sigma_{(q \bar q)^{J=1}_{L,T} p} (r^\prime_\bot, W^2) 
\nonumber \\
&&=  \frac{6 \alpha}{2 \pi^2} Q^2 \Sigma_q Q^2_q 
     \left\{ \matrix{
       4 \int dzz (1-z) \int d^2 r^\prime_\bot K^2_0 (r^\prime_\bot Q)
       \sigma_{(q \bar q)^{J=1}_L p} (r^\prime_\bot, W^2), \cr
        ~~~~~~~\cr
       \int dz (1-2z (1-z)) \int d^2 r^\prime_\bot K^2_1 (r^\prime_\bot Q)
      \sigma_{(q \bar q)^{J=1}_T p} (r^\prime_\bot, W^2). }
    \right.
\label{19}
\eqa
In (\ref{19}), $\Sigma_q Q^2_q$ denotes the sum over the quark charges 
(with $3 
\Sigma Q^2_q \equiv R_{e^+e^-}$), and $K_0 (r^\prime_\bot Q)$ and $K_1 
(r^\prime_\bot Q)$ are modified Bessel functions. The essential point in
(\ref{19}), in distinction from the usually employed \cite{Golec} form
of the dipole picture\footnote{Compare \cite{Ewerz} for a
careful examination of the usual formulation of the CDP.}, is the
factorization of the $z(1-z)$ dependence identical in form to the 
(squares of the ) longitudinal and transverse electromagnetic-current 
components in (\ref{7}) and (\ref{8}). 
Indeed, (\ref{19}) follows from the formulation of the CDP
in terms of the interquark transverse separateion, $\vec r_\perp$
\bqa
&\sigma_{\gamma^*_{L,T}p} (W^2, Q^2) = \frac{6 \alpha}{2 \pi^2}
Q^2 \Sigma_q Q^2_q \cdot \nonumber\\
&\cdot \left\{ \begin{array}{l@{}l}
& 4 \int dz z^2(1-z)^2 \int d^2 r_\bot K^2_0 (r_\bot \sqrt{z(1-z)}Q)
\sigma_{(q \bar q)p} (r_\bot, z(1-z),W^2), \\
& \int dz (1-2z (1-z)) z (1-z) \int d^2 r_\bot K^2_1 (r_\bot \sqrt{z(1-z)})
\sigma_{(q \bar q)p} (r_\bot, z(1-z),W^2),
\end{array} \right. 
\label{20}
\eqa
by requiring a $z(1-z)$ dependence identical in form to the
one in (\ref{7}) and (\ref{8}),
\bqa
  &&\int dz z(1-z)\sigma_{(q\bar q)p}\bigl({{r_\perp^\prime}\over{\sqrt{z(1-z)}}},
      z(1-z), W^2\bigr)  \nonumber \\
  &=& \int dz z(1-z)\sigma_{(q\bar q)_L^{j=1}}\bigl(r_\perp^\prime, W^2\bigr),
\label{21} 
\eqa
and 
\bqa
  && \int dz \bigl(1-2z(1-z)\bigr)\sigma_{(q\bar q)p}
    \bigl({{r_\perp^\prime}\over{\sqrt{z(1-z)}}}, z(1-z), W^2\bigr) 
    \nonumber \\
  &=& \int dz \bigl(1-2z(1-z)\bigr)\sigma_{(q\bar q)_T^{j=1}}
    \bigl(r_\perp^\prime, W^2\bigr).
\label{22} 
\eqa
Substitution of $\vec r_\perp$ in terms of $\vec r_\perp^{~\prime}$ 
in (\ref{20}) and substitution of 
(\ref{21}) and (\ref{22}) immediately imply (\ref{19}).

The factorization implies the
discrimination between dipole cross sections for longitudinally and
transversely polarized $(q \bar q)^{J=1}$ states in (\ref{19}).
The form (\ref{19}) of the color-dipole picture (CDP) will be referred
to as the $r^\prime_\bot$ representation. 

Carrying out the integration over $z$ in (\ref{19}), we have
\be
\sigma_{\gamma^*_{L,T}} (W^2, Q^2) = \frac{2 \alpha R_{e^+e^-}}{3 \pi^2}
Q^2 \int d^2 r^\prime_\bot K^2_{0,1} (r_\bot^\prime Q) 
\sigma_{(q \bar q)^{J=1}_{L,T} p}
(r^\prime_\bot, W^2).
\label{23}
\ee
In (\ref{19}) and (\ref{23}), we now introduce the $q \bar q$-size
effect (\ref{17}) that becomes

\be
    \sigma_{(q \bar q)^{J=1}_T p} (r^\prime_\bot, W^2) = \rho 
    \sigma_{(q \bar q)^{J=1}_L p} (r^\prime_\bot, W). 
   \label{24}
\ee

To incorporate the coupling of the $q \bar q$ color dipole to two
gluons, the $r^\prime_\bot$ representation must be supplemented by

\be
    \sigma_{(q \bar q)^{J=1}_T p} (r^\prime_\bot, W^2) = 
\rho \sigma_{(q \bar q)^{J=1}_L p} (r^\prime_\bot, W^2) =
\rho \int d^2 l^\prime_\bot \bar \sigma_{(q \bar q)^{J=1}_L p}
(\vec l^{~\prime 2}_\bot, W^2)    
(1-e^{-i \vec l^{~\prime}_\bot \cdot \vec r^{~\prime}_\bot}),
\label{25}
\ee
where (\ref{24}) was incorporated. In (\ref{25}), $\vec l^{~\prime}_\bot$,
is related to the transverse momentum of the gluon, $\vec l_\bot$,
absorbed by the quark or antiquark by
\be
   \vec l^{~\prime}_\bot = \frac{\vec l_\bot}{\sqrt{z(1-z)}}.
\label{26}
\ee
In the $r^{\prime 2}_\bot \to 0$ limit, (\ref{25}) may be approximated by

\be
   \sigma_{(q \bar q)^{J=1}_T p} (r^\prime_\bot, W^2) 
= \rho \sigma_{(q \bar q)^{J=1}_L p} (r^\prime_\bot, W^2)
\cong \rho \vec r^{~\prime 2}_\bot \frac{\pi}{4} \int d
\vec l^{~\prime 2}_\bot \vec l^{~\prime 2}_\bot
\bar \sigma_{(q \bar q)^{J=1}_L} (\vec l^{~\prime 2}_\bot, W^2).
     \label{27}
\ee
The generic structure of two-gluon couplings to the $q \bar q$-pair
implies ``color transparency'' \cite{Nikolaev}: vanishing of the
color-dipole interaction for vanishing interquark distance. Due
to the strong decrease of the modified Bessel functions $K_0 
(r^\prime_\bot Q)$ and $K_1 (r^\prime_\bot Q)$ in (\ref{23}) for large
values of their argument, $r^\prime_\bot Q$, the large-$Q^2$ behavior
of $\sigma_{\gamma^*_{L,T}p} (W^2, Q^2)$ in (\ref{23}) can be obtained
by substitution of the small-$r^\prime_\bot$ approximation (\ref{27}).

In passing, we quote the explicit ansatz \cite{Surrow} consistent with
(\ref{27}) that was previously used in a (successful) representation
of the experimental data\footnote{Actually $\rho = 1$, i.e. helicity 
independence was used in the description of 
the experimental data for $\sigma_{\gamma^*p} (W^2, Q^2)$. The difference
of $\rho = 1$ and $\rho = 4/3$ from (\ref{15}) is mainly relevant for
$R(W^2, Q^2)$.}, 
\be
   \sigma_{(q \bar q)^{J=1}_T p} (r^\prime_\bot, W^2) 
= \rho \sigma_{(q \bar q)^{J=1}_L} (r^\prime_\bot, W^2)
= \rho \sigma^{(\infty)} (W^2) (1 - J_0 (r^\prime_\bot
    \Lambda_{sat}^2(W^2))),
\label{28}
\ee
with a power-law ansatz for the ``saturation scale'' $\Lambda^2_{sat} (W^2)$,
\bq
    \Lambda_{sat}^2(W^2) ={\pi\over{\sigma^{(\infty)}(W^2)}}  
   \int d l^{\prime 2}_\bot l^{\prime 2}_\bot \bar 
\sigma_{(q \bar q)^{J=1}_L p}
    ( l^{\prime 2}_\bot, W^2).
\label{29}
\eq
The hadronic cross section, $\sigma^{(\infty)} (W^2)$, is approximately
constant. The ansatz (\ref{28}) is mentioned in order to explicitly display 
the role of $\rho$ as a factor that determines the relative magnitude of
the total hadronic cross sections for transversely relative to
longitudinally polarized $(q \bar q)^{J=1}$ states; the hadronic cross
section, $\sigma^{(\infty)} (W^2)$, for scattering of $(q \bar q)^{J=1}_L$
states of mass $M_{q \bar q}$ is effectively replaced by $\rho \sigma^{(\infty)} (W^2)$
when passing from longitudinally polarized to transversely polarized
$(q \bar q)^{J=1}$ states, compare the discussion following (\ref{17}).

A unique consequence on the ratio $R (W^2, Q^2)$ of the 
longitudinal to the transverse photoabsorption cross section at large values
of $Q^2$ follows immediately by substituting the $r^{\prime 2}_\bot \to 0$
approximation (\ref{27}) into the $r^\prime_\bot$-representation of the
CDP (\ref{23}). Indeed,
in the large-$Q^2$ limit, the dependence on the details of the $(q \bar q)p$
interaction, compare (\ref{28}) as an example, cancels in 
$R(W^2, Q^2)$, and, for $Q^2$ sufficiently large,
we have
\be
R(W^2, Q^2) \equiv \frac{\sigma_{\gamma^*_L p}(W^2, Q^2)}{\sigma_{\gamma^*_T p}
 (W^2, Q^2)} =
\frac{\int d^2 r^\prime_\bot r^{\prime 2}_\bot K^2_0 (r^\prime_\bot Q)}
{\rho \int d^2 r^\prime_\bot r^{\prime 2}_\bot K^2_1 (r^\prime_\bot Q)} = 
\frac{1}{2\rho} = \frac{3}{8} = 0.375.
\label{30}
\ee
Equivalently, in terms of the structure functions,
\be
     \frac{F_L (W^2, Q^2)}{F_2 (W^2, Q^2)} = \frac{1}{1+2 \rho} = 
     \frac{3}{11} \simeq 0.27.
\label{31}
\ee
In (\ref{30}) and (\ref{31}), the equality \cite{Gradsteyn} 
\be
\int^\infty_0 dy~y^3 K^2_0 (y) = \frac{1}{2} \int^\infty_0 dy~y^3 K^2_1 (y)
\label{32}
\ee
was used, and the value of $\rho = 4/3$ from (\ref{15}) was inserted. We note 
that helicity independence, $\rho = 1$, leads to $R (W^2, Q^2) = 0.5$
and $F_L/F_2 = 1/3$. 
This case of $\rho = 1$, using the ansatz (\ref{28}), in ref. \cite{Tentyukov}
was evaluated and discussed in detail, including the transition to 
$Q^2 \to 0$, and compared with the H1 analysis  of the longitudinal 
structure function available at the time\cite{H1}
 that was based on certain
theoretical input assumptions.
Compare fig.1.  Replacing $\rho=1$ by $\rho={4\over 3}$ decreases
the theoretical predictions for $\sigma_{\gamma^*_Lp}(W^2,Q^2)$ in fig.1 
by the factor of ${9\over {11}}\cong 0.82$,
improving agreement with the data.

We stress that the predictions (\ref{30}) and
(\ref{31}) only rely on the CDP in the $r^\prime_\bot$ representation given
by (\ref{19}), (\ref{23})  combined with color transparency
(\ref{27}) and the $q \bar q$-transverse-size effect incorporated into the
proportionality (\ref{24}).
The predictions  are independent  of any specific ansatz for the 
dipole-cross section.
We note that the predictions (\ref{30}) and (\ref{31}) are most reliable for
$5 GeV^2 \lsim Q^2 \lsim 100 GeV^2$. They become less reliable, when 
$Q^2$ increases to $Q^2 \gg 100 GeV^2$, since in this case the CDP has to
be refined by a restriction on the actively contributing $q \bar q$ masses,
$M_{q \bar q}$ \cite{PRD}.
For a value of $F_2 \cong 1.2$,
that is typical for the $Q^2$ and $W$ values where final separation data will
become available\footnote{Preliminary results \cite{Antunovic} from HERA
were presented at DIS 2008, 7-11 April 2008, University College London}, 
according to (\ref{31}), we find $F_L \simeq 0.32$.

The parameter $\rho$ from (\ref{24}), making use of the first equality in
(\ref{31}), can be determined from measurements of DIS at different 
electron-proton center-of-mass energies, $\sqrt s$, for fixed values of
$x$ and $Q^2$. The reduced cross section of DIS is given by\footnote{Compare
e.g. ref. \cite{Devenish}}
\be
\sigma_r (x, y, Q^2) = F_2 (x, Q^2) 
\left(1 - \frac{y^2}{1 + (1-y)^2} \frac{1}{1 + 2 \rho} \right),
\label{33}
\ee
where $y = Q^2/xs$. The slope of a straight-line fit of
$\sigma_r (x,y,Q^2)$ as a function of $0 \le y^2/\left(1+(1-y)^2 \right)
\le 1$ determines $\rho$. A value of 
\be
\rho = 1
\label{34}
\ee
corresponds to helicity independence, i.e. equality of the forward scattering
amplitudes of $(q \bar q)^{J=1}_h$ fluctuations of the photon on the proton
for helicities $h = 0,~ h = +1$ and $h = -1$. A deviation from $\rho = 1$
rules out helicity independence. Longitudinally polarized $(q \bar q)^{J=1}$
states have a reduced transverse size relative to transversely polarized
$(q \bar q)^{J=1}$ states. This implies a deviation from $\rho = 1$ of
predictible magnitude, compare (\ref{16}). The theoretically preferred
value of $\rho$, accordingly, 
is\footnote{In refs. \cite{ks3} and \cite{DIS07},
the interpretation of $\rho$ in terms of a proportionality of sea quark and
gluon distributions led to a certain theoretical preference of (\ref{34})
versus (\ref{35}) in contrast to our present point of view. This needs some
further investigation.}
\be
\frac{1}{\rho} = \frac{3}{4}.
\label{35}
\ee
The measurement of $\rho$ provides insight into the dynamics of the scattering
of the $q \bar q$ fluctuations of the photon on the proton and a strong
constraint on the CDP.

An interesting upper bound on $R (W^2, Q^2)$ was recently derived 
in the framework of the CDP. The bound is given by \cite{Ewerz}
\be
     R(W^2, Q^2) \le 0.37248,
\label{36}
\ee
or, in terms of $\rho$ with $R(W^2, Q^2) = 1/2 \rho$ from (\ref{30}),
$\rho>1.34235$.  The approximate numerical coincidence of our 
prediction (\ref{30}) with the upper bound (\ref{36}) is accidental.

The upper bound (\ref{36}) is obtained from (\ref{20}) by adopting
the frequently employed approximation of $z(1-z)$ independence of the 
ansatz for the color-dipole cross section
\bq
   \sigma_{(q\bar q)p}(r_\perp,z(1-z),W^2) \equiv 
   \sigma_{(q\bar q)p}(r_\perp,W^2)
\label{37})
\eq
In this case of (\ref{37}), the integration over $z$ may be
carried out in (\ref{20}), and the longitudinal and transverse
cross sections become integrals over $r_\perp$-dependent 
probability densities multiplied by the dipole cross section (\ref{37}).
The bound (\ref{36}) follows from the maximum of the ratio
of the longitudinal-to-transverse probability densities as a function of $\vec r_\perp$.
If the assumption (\ref{37}) is dropped, the derivation of the bound
fails,

The bound (\ref{36}) on $R(W^2,Q^2)$ means that models for the dipole
cross section that imply violations of the bound must necessarily 
contain a dependence on the configuration variable $z(1-z)$.  
An example is provided by our ansatz (\ref{28})\cite{Surrow} with $\rho=1$ and
$R(W^2,Q^2)=0.5$.  Experimental values of $R(W^2,Q^2)$ below the bound
(\ref{36}) neither require nor rule out a $z(1-z)$ dependence
of the dipole cross sections.  Indeed, the theoretical restriction (\ref{3}) 
on the contributing $q\bar q$ masses even requires
the approximation (\ref{37}) to break down
at a certain level of accuracy in dependence on the range of the
kinematical variables $Q^2$ and $W^2$.

The $r^\prime_\bot$ representation rests on an explicit factorization of
the photoproduction cross section in terms of a three-step process:
$\gamma^* (q \bar q)$ coupling, $(q \bar q)$ propagation and $(q \bar q)$
scattering.  The factorization is intimately related to
the underlying notion of a $q \bar q$ fluctuation of the photon interacting
with the proton. It even appears as an unavoidable consequence of this
picture. The transverse size of the $(q \bar q)^{J=1}$ state emerging
from the photon being large for transverse relative to longitudinal
polarization, we predict $R (W^2, Q^2)= 3/8 = 0.375$ or, equivalently,
$F_L (W^2, Q^2)/F_2 (W^2, Q^2) \cong 0.27$.

\vspace{1 cm}
\noindent{\bf Acknowledgement}\\
 Useful discussions with Allen Caldwell and Vladimir Chekelyan
on the Hera experimental data are gratefully acknowledged.

\vfill\eject

\vspace{1 cm}
\noindent{Figure caption}\\
Fig.1: ~The longitudinal photoabsorption cross section 
$\sigma_{\gamma^*_Lp}(W^2,Q^2)\equiv \sigma_L$ from ref.\cite{Tentyukov}
as a function of the scaling variable $\eta\equiv 
(Q^2+m_0^2)/\Lambda_{sat}^2(W^2)$
compared with HERA data available in 2001 and based on an H1  
analysis\cite{H1} with theoretical QCD input assumptions rather than 
on a longitudinal-transverse separation measurement.


\begin{figure}[htbp]\centering
\epsfysize=10cm
\centering{\epsffile{fig1.eps}}
\label{fig1}
\end{figure}

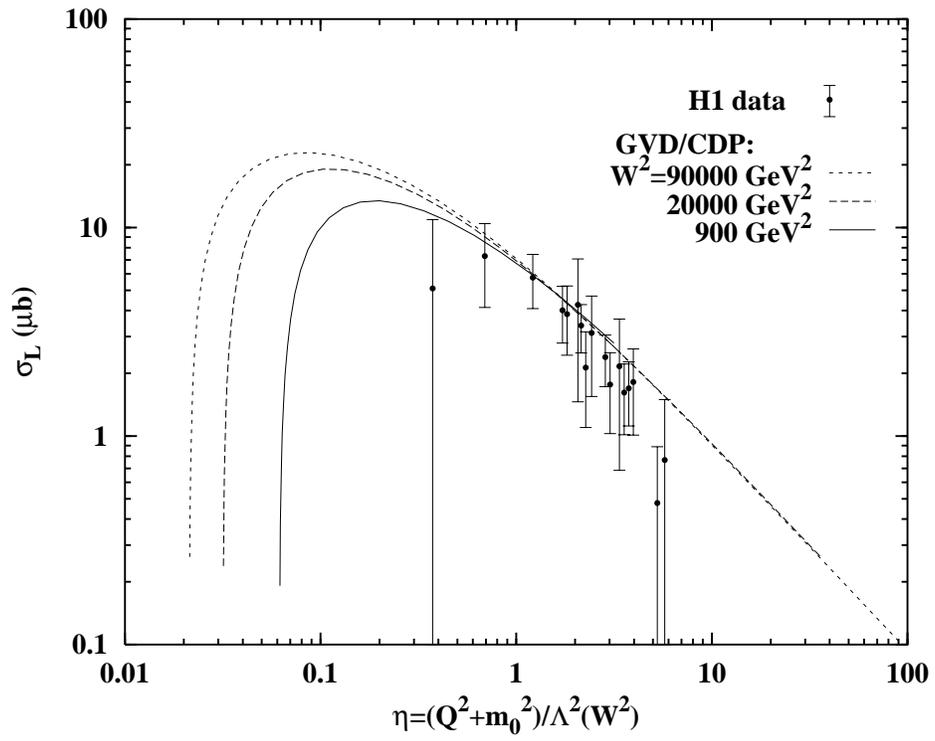
\end{document}